\documentclass[aps,prb,floatfix,amsmath,amssymb,eqsecnum,nofootinbib,twocolumn,showpacs]{revtex4}
\usepackage{graphicx}
\usepackage{dcolumn}
\usepackage{bm}
\usepackage{epstopdf}
\usepackage[bookmarks=true,colorlinks,linkcolor=blue,urlcolor=blue,citecolor=blue]{hyperref}
\usepackage{color}
\usepackage[section]{placeins}
\usepackage{float}

\newcommand{\expect}[1]{\langle {#1} \rangle}

\usepackage{setspace} 
\newcommand{\beq}{\begin{equation}}
\newcommand{\eeq}{\end{equation}}
\newcommand{\bea}{\begin{eqnarray}}
\newcommand{\eea}{\end{eqnarray}}
\newcommand{\nn}{\nonumber\\}

\usepackage{amsmath}
\usepackage{amssymb}

\newcommand{\bk}{{\bm k}}

\newcommand{\bK}{{\bm K}}

\newcommand{\bH}{{\bm H}}
\newcommand{\br}{{\bm r}}

\begin{document}

\title{Mean field theory of competing orders\\ in metals with antiferromagnetic exchange interactions}

\author{Jay Deep Sau}
\affiliation{Department of Physics, University of Maryland, College Park MD}

\author{Subir Sachdev}
\affiliation{Department of Physics, Harvard University, Cambridge MA
02138}

\date{\today }

\begin{abstract}
It has long been known that two-dimensional metals with antiferromagnetic exchange interactions have a weak-coupling instability
to the superconductivity of spin-singlet, $d$-wave electron pairs. We examine additional possible instabilities in the spin-singlet particle-hole channel,
and study their interplay with superconductivity. We perform an unrestricted Hartree-Fock-BCS analysis of bond order parameters in 
a single band model on the square lattice with nearest-neighbor exchange and repulsion, while neglecting on-site interactions.
The dominant particle-hole instability is found to be an incommensurate, bi-directional, bond density wave with wavevectors along the $(1,1)$
and $(1,-1)$ directions, and an internal $d$-wave symmetry. The magnitude of the ordering wavevector is close to the separation between points on the
Fermi surface which intersect the antiferromagnetic Brillouin zone boundary. The temperature dependence of the superconducting and bond order parameters
demonstrates their mutual competition. We also obtain the spatial dependence of the two orders in a vortex lattice induced by an applied magnetic field:
``halos'' of the bond order appear around the cores of the vortices. 
\end{abstract}


\maketitle

\section{Introduction}

All of the quasi-two dimensional higher temperature superconductors are proximate to metals with strong local
antiferromagnetic exchange interactions \cite{chubukov}. In some cases, there is even a antiferromagnetic quantum critical point,
with a diverging spin correlation length, in the region of the highest critical temperatures for superconductivity \cite{matsuda,matsuda2}. 
However, in the materials with highest critical temperatures, the hole-doped cuprates, the regions with the strongest superconductivity
are well separated from the antiferromagnetic quantum critical point \cite{julien2}. Interestingly, it is in these same materials that the 
`pseudogap' regime is best defined, along with the presence of competing charge density wave orders \cite{julienvortex,keimer,chang,hawthorn,o6}.

In the context of a weak-coupling treatment of the antiferromagnetic exchange interactions \cite{sc1,sc2,sc3,sc4,sc5,sc6}, it has long been known that
unconventional spin-singlet superconductivity can appear, with a gap function which changes sign between regions of the Fermi surface
connected by the antiferromagnetic ordering wavevector (this corresponds to $d$-wave pairing, in the context of the cuprate 
Fermi surface). Recently, quantum Monte Carlo simulations have shown \cite{sdwsign} that this mechanism of superconductivity via exchange of antiferromagnetic fluctuations survives in the strongly-coupled regime across the quantum critical point.

Field-theoretical studies \cite{advances,metlitski10-2,metlitski-njp} 
of the vicinity of the antiferromagnetic quantum critical point in a two-dimensional metal have also 
found strong evidence for the dominance of a $d$-wave superconducting instability. These studies \cite{metlitski10-2,metlitski-njp,pepin} 
also noted that an instability to a particular
type of charge order, an incommensurate $d$-wave bond order, was nearly degenerate with the superconducting instability.
These results suggest that a combination of fluctuating superconducting and charge orders could describe the pseudogap
regime of the hole-doped cuprate superconductors; Ref.~\onlinecite{o6} has shown that a theory of these fluctuating orders 
describes the X-ray scattering data\cite{keimer,chang,hawthorn} well. 
However, to establish such a proposal theoretically, we need to understand the evolution
of these instabilities in a metal which is well separated from the antiferromagnetic quantum critical point.

\begin{figure}
\centering
\includegraphics[width=2.8in]{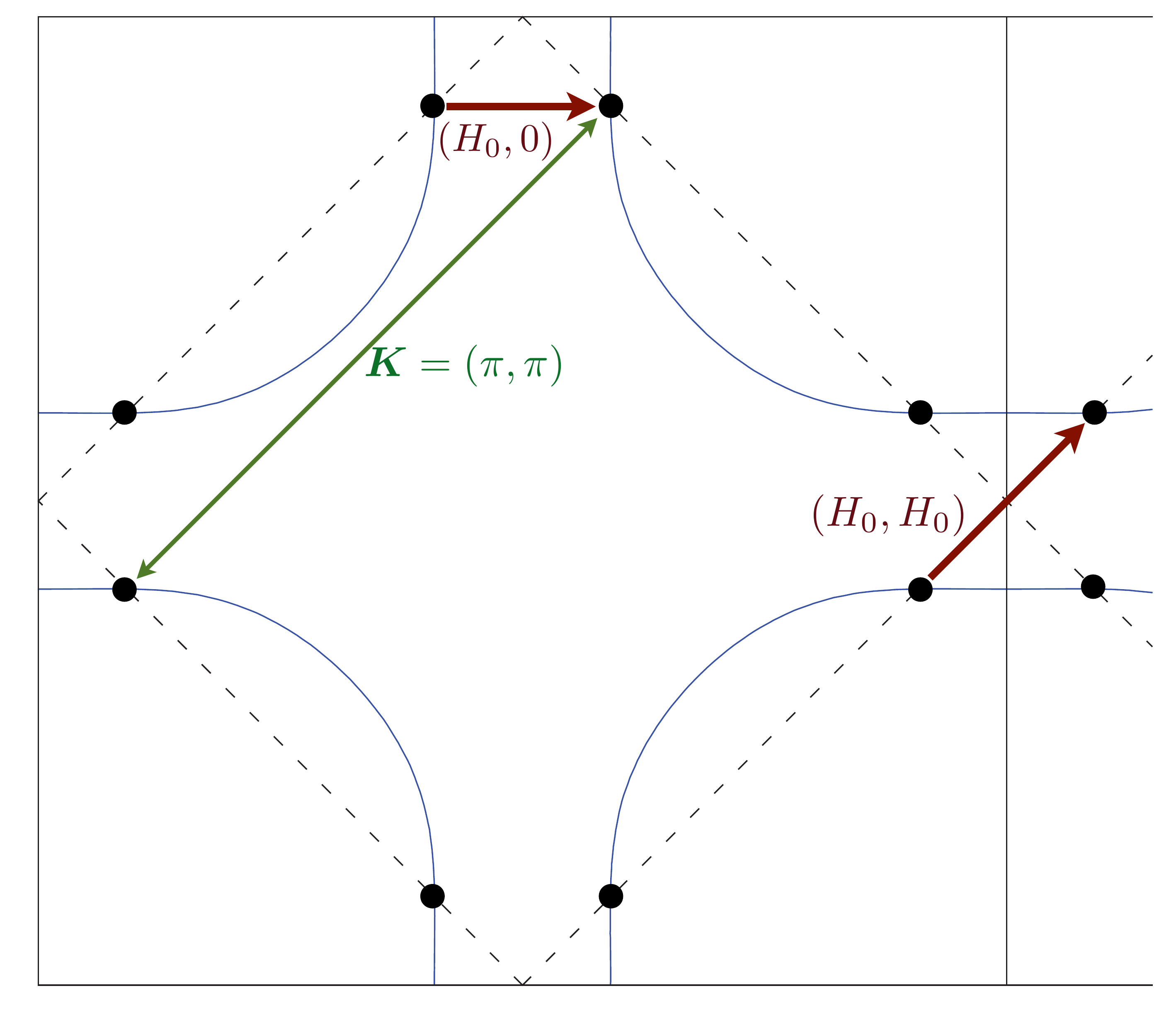}
\caption{Fermi surface in the square lattice Brillouin zone, showing the hot spots (filled circles), and defining the value of $H_0$.}
\label{fig:bz}
\end{figure}

A linear stability analysis of a two-dimensional metal with short-range antiferromagnetic interactions, but away from the antiferromagnetic
quantum critical point, was carried
out recently in Ref.~\onlinecite{rolando} (related studies are in Refs.~\onlinecite{metzner1,metzner2,yamase,kee,kampf,laughlin,seamus}). Ref.~\onlinecite{rolando} found that the leading instability upon cooling down from high temperatures was to 
a $d$-wave superconductor. However, if one ignored this instability, and looked for the leading instability in the particle-hole channel,
it was found to be an incommensurate charge density wave with a $d$-wave form factor, and wavevectors along the $(1,\pm 1)$ directions
of the square lattice. Moreover, the optimum wavevectors of the instability, $(H, \pm H)$ were very close to the separation $(H_0, \pm H_0)$
between certain
``hot spots'' on the Fermi surface (see Fig.~\ref{fig:bz}); leading subdominant saddle points were also found close to the wavevectors
$(\pm H_0, 0)$, $(0, \pm H_0)$ which are observed in the X-ray scattering.
The hot spots are special points on the Fermi surface which are separated
from one other hot spot by the antiferromagnetic wavevector $\bK = (\pi, \pi)$. The wavevector magnitude $H_0$ appeared in the 
results,\cite{rolando,metzner1}
even though they were not treated as special Fermi surface points in the computation, and the antiferromagnetic correlations were short-ranged.
But $\bK$ can nevertheless  be expected to play a role because
the Fourier transform of the antiferromagnetic exchange interaction has an extremum at or near $\bK$; in Ref.~\onlinecite{metzner1}, the
ordering wavevectors were related to the crossing points of $2k_F$ singularities. 
These results were 
interpreted as evidence for the applicability of the theory of competing and nearly degenerate superconducting and charge-density
wave orders away from the antiferromagnetic quantum critical point.

In this paper, we will extend the earlier computation\cite{rolando} 
beyond the regime of linear instability at high temperature, to a complete determination
of the optimal state at low temperature, with co-existing superconducting and charge-density-wave order parameters.
This study is the analog of that in Ref.~\onlinecite{yamase2} for the case of competition of superconductivity with Ising-nematic
order at zero wavevector.
We will also study the solution in the presence of an applied magnetic field and a vortex lattice.

We begin in Section~\ref{sec:mft} by setting up the Hartree-Fock-BCS equations for the square lattice model
with nearest-neighbor exchange interactions ($J$) and a nearest-neighbor Coulomb repulsion ($V$). All the variational 
parameters of the mean-field theory will lie on the nearest-neighbor links: these include a spin-singlet electron pairing
amplitude, $Q_{ij}$, and a spin-singlet particle-hole pairing amplitude, $P_{ij}$. We will allow for arbitrary spatial
dependencies in the $P_{ij}$ and $Q_{ij}$, including the possibility of time-reversal symmetry breaking solutions with
non-zero currents on the links. However, our analysis does not include on-site interactions or on-site variational parameters:
we discuss shortcomings in experimental applications possibly due to this omission in Section~\ref{sec:conc}.

Our main results on the numerical solutions of the mean-field equations appear in Section~\ref{sec:num}.
We always find that the dominant instability in the particle-hole channel is a bi-directional bond density wave
with wavevectors very close to the diagonal values $(\pm H_0, \pm H_0)$ determined by the intersections between
the Fermi surface and antiferromagnetic Brillouin zone boundary (Fig~\ref{fig:bz}. The bond density wave can co-exist 
with $d$-wave superconductivity, and we obtain results on the temperature dependence of the two orders. 
We also present a solution in the presence of an applied magnetic field, displaying vortices in the superconducting order
surrounded by ``halos'' of bond order, similar to the initial observations by Hoffman {\em et al.}\cite{jenny}
We note that there have been earlier Hartree-Fock-BCS studies of competing
orders around vortex cores\cite{ting,martin}, but they had local antiferromagnetic order as the driving mechanism.

In Section~\ref{sec:hotspot} we introduce a simplified momentum space model which has instabilities similar to those
obtained in the full lattice model in Section~\ref{sec:num}. This provides a simple and useful physical picture of the 
structure of the phase diagram. However, it is not possible to extend this model to the spatially inhomogeneous case that
arises in the presence of an applied magnetic field.

Section~\ref{sec:conc} will discuss missing ingredients in our model computations which could possibly yield
charge order at the observed wavevectors $(\pm H_0, 0)$, $(0, \pm H_0)$.

We note that a similar Hartree-Fock-BCS computation on a $t$-$J$ model has been carried out recently by Laughlin\cite{laughlin},
and he finds a dominant instability in the particle-hole channel of a state with staggered orbital currents\cite{marston,kotliar,sudip,leewen,laughlin}.
However, his computation does not consider Fermi surfaces with hot spots
as in Fig.~\ref{fig:bz}. Our computations do include the staggered orbital current state as a possible saddle point, and for our chosen parameters,
the bond density waves described below have a lower energy.
Similarly, we also allowed for states with uniform orbital currents,\cite{varma} and did not find them.

\section{Model and mean field equations}
\label{sec:mft}

We will examine the $t$-$J$-$V$ on the square lattice, with the dispersion chosen to ensure that there are 
hot spots on the Fermi surface:
\bea
H &=& H_t + H_{JV} \nn
H_t &=& \sum_{\bk, \alpha} \varepsilon (\bk) c_{\bk ,\alpha}^\dagger c_{\bk, \alpha} \nn
H_{JV} &=& \sum_{i<j} \left[ \frac{J_{ij}}{4} \sigma^a_{\alpha\beta} \sigma^{a}_{\gamma\delta}\label{eq:H}
c^\dagger_{i\alpha} c_{i \beta} c^\dagger_{j \gamma} c_{j \delta} \right. \nn
&~& \quad\quad \quad \left. + V_{ij} c_{i\alpha}^\dagger c_{i \alpha} c_{j \beta}^\dagger c_{j \beta} \right].
\eea
The dispersion is that in Ref.~\onlinecite{rolando}, with
\bea
\varepsilon (\bk ) &=& - 2 t_1 \left( \cos (k_x) + \cos (k_y) \right) - 4 t_2 \cos (k_x) \cos (k_y) \nn
&~& - 2 t_3 \left( \cos (2 k_x) + \cos (2 k_y) \right) - \mu
\eea
where $t_1 = 1$, $t_2 = -0.32$, and $t_3 = -0.5 t_2$. 

It is useful to set up the Hartree-Fock-BCS mean field equations by working in real space. We will work with the hypothesis
that competing orders are controlled by the bond expectation values \cite{ssrmp}, and neglect on-site factorizations of the 4 fermion terms.
Then, for each pair of sites, $i,j$, for which the $J_{ij}$ or $V_{ij}$ are non-zero, we have  a pair of complex numbers which will serve
as the variational parameters of our mean-field theory; we define these by
We define the bond expectation values
\bea
\left\langle c^{\dagger}_{i \alpha} c^\dagger_{j \beta} \right\rangle &=& Q_{ij} \varepsilon_{\alpha\beta} \nn
\left\langle c_{i \alpha} c_{j \beta} \right\rangle &=& -Q^\ast_{ij} \varepsilon_{\alpha\beta} \nn
\left\langle c^{\dagger}_{i \alpha} c_{j \beta} \right\rangle &=& P_{ij}  \delta_{\alpha\beta} 
\eea
with $Q_{ij} = Q_{ji}$ and $P_{ji} = P_{ij}^\ast$. With these definitions, the
Hartree-Fock-BCS factorization of $H$ is
\bea
H_{HF} = \sum_{i,j} (c_{i \uparrow}^{\dagger}, c_{i \downarrow} ) \left( 
\begin{array}{cc} A_{ij} & B_{ij} \\
B^\ast_{ij} & - A_{ji} 
\end{array} \right) \left( \begin{array}{c} c_{j \uparrow} \\ c_{j \downarrow}^\dagger \end{array} \right) 
\eea
where
\bea
A_{ij} &=& - t_{ij} - \left( 3J_{ij}/4 + V_{ij} \right) P^\ast_{ij} - \delta_{ij} \mu  \nn
B_{ij} &=& - \left( 3J_{ij}/4 - V_{ij} \right) Q^\ast_{ij} 
\eea
$H_{HF}$ has eigenvalues which occur in pairs, $\omega_\mu, - \omega_\mu$ ($\omega_\mu > 0$), 
and we write the corresponding eigenvectors as 
\beq
\left( \begin{array}{c} \mathcal{U}_{i \mu} \\ \mathcal{V}_{i \mu} \end{array} \right) , \left( \begin{array}{c} -\mathcal{V}^\ast_{i \mu} \\ \mathcal{U}^\ast_{i \mu} \end{array} \right).\label{eq:ev}
\eeq

Orbital effects of the magnetic field can be introduced by minimal substitution into the hopping parameters as 
\beq
t_{ij}=t_{ij}^{(0)}e^{i a_{ij}},\label{eq:t}
\eeq
where $t_{ij}^{(0)}$ are hoppings in the absence of a magnetic field and $a_{ij}=\int_{\bm r_i}^{\bm r_j}d^3\bm r\cdot \bm a(\bm r)$ 
are Aharonov-Bohm phases resulting from the vector potential $\bm a(\bm r)$ generated by the magnetic field. While the magnetic field does not 
break translation invariance in the strict sense, $H_{HF}$ can be translationally invariant in a specific gauge only in a magnetic unit cell, 
which encloses a full electron (and therefore two superconducting) flux quantum(quanta). In order to maintain spatial translational invariance 
of the vector potential we insert a pair of superconducting flux quanta (i.e. containing flux $h/2 e$) in a pair of plaquettes 
of the square lattice to compensate the flux of the uniform magnetic field. At the same time we introduce a $\pi$ phase-shift branch cut  
for the electrons joining the two plaquettes.  The magnetic field leads the superconducting 
order parameter $Q_{ij}$ to develop a vortex lattice structure. The specific profile for the vector potential described, 
whose precise description is provided in Appendix A, is chosen so as   to simplify the numerical stability of the self-consistency procedure, 
following previous work \cite{vafek}.
Specifically, in this gauge the phase of the self-consistent order parameter is expected to be nearly constant so that the initial guess of a 
constant order parameter would have a phase profile close to the final answer. 
One would still have a non-vanishing supercurrent proportional to the non-vanishing vector potential $\bm a$ i.e. 
 $\bm a|Q|^2$ term in the current density. In our calculations, the vortex cores, which appear as dips in the SC order parameter, 
are found to exist near the superconducting flux quanta.
The details of the calculation of the vector potential $\bm a$ is given in Appendix A.

Introducing the Bogoliubov operators $\gamma_{\mu\alpha}$ we then have
\beq
H_{HF} = \sum_{\mu} \omega_\mu \left( \gamma_{\mu\alpha}^\dagger \gamma_{\mu\alpha} - 1 \right)
\eeq
where the unitary transformation to the Bogoliubov operators is
\beq
c_{i \alpha} = \sum_{\mu} \left( \mathcal{U}_{i \mu} \gamma_{\mu \alpha} - \mathcal{V}_{i \mu}^\ast \varepsilon_{\alpha\beta} \gamma_{\mu \beta}^\dagger \right).
\eeq
So we have the expectation values
\bea
&& \left\langle c^{\dagger}_{i \alpha} c^\dagger_{j \beta} \right\rangle_{HF} = \widetilde{Q}_{ij} \varepsilon_{\alpha\beta} \nn
&&= -\varepsilon_{\alpha\beta} \sum_\mu \left( \mathcal{U}_{i \mu}^\ast \mathcal{V}_{j \mu} + \mathcal{U}_{j \mu}^\ast \mathcal{V}_{i \mu} \right) \frac{1}{2} \tanh\left( \frac{\omega_\mu}{2T} \right)\nn
&& \left\langle c_{i \alpha} c_{j \beta} \right\rangle_{HF} = -\widetilde{Q}^\ast_{ij} \varepsilon_{\alpha\beta} \nn
&& \left\langle c^{\dagger}_{i \alpha} c_{j \beta} \right\rangle_{HF} = \widetilde{P}_{ij}  \delta_{\alpha\beta} \\
&&= -\delta_{\alpha\beta} \sum_\mu \left( \mathcal{U}_{i \mu}^\ast \mathcal{U}_{j \mu} - \mathcal{V}_{j \mu}^\ast \mathcal{V}_{i \mu} \right) \frac{1}{2} \tanh\left( \frac{\omega_\mu}{2T} \right), \nonumber
\eea
which defines the complex numbers $\widetilde{Q}_{ij}$ and $\widetilde{P}_{ij}$ as functions of the $Q_{ij}$ and $P_{ij}$.
Above, we used the orthogonality relations
\bea
\sum_{\mu} \left( \mathcal{U}_{i \mu}^\ast \mathcal{U}_{j \mu} + \mathcal{V}_{j \mu}^\ast \mathcal{V}_{i \mu} \right) &=& \delta_{ij} \nn
\sum_{\mu} \left( \mathcal{U}_{i \mu}^\ast \mathcal{V}_{j \mu} - \mathcal{U}_{j \mu}^\ast \mathcal{V}_{i \mu} \right) &=& 0.
\eea
Finally, the variational estimate for the free energy is
\bea
F &=& F_{HF} + \left\langle H - H_{HF} \right\rangle \nn
&=& \sum_{\mu} \left( - \omega_\mu - 2 T \ln ( 1 + e^{-\omega_\mu/T} ) \right) \label{eq:F}\\
&+& \sum_{i<j} \Biggl[ 2 \left(\frac{3J_{ij}}{4}  - V_{ij} \right) \left[- | \widetilde{Q}_{ij} |^2 + Q_{ij} \widetilde{Q}_{ij}^\ast + Q_{ij}^\ast \widetilde{Q}_{ij} \right]  \nn
&+& 2 \left(\frac{3J_{ij}}{4}  + V_{ij} \right) \left[- | \widetilde{P}_{ij} |^2 + P_{ij} \widetilde{P}_{ij}^\ast + P_{ij}^\ast \widetilde{P}_{ij} \right] 
\Biggr]. \nonumber
\eea
Our task is to minimize this free energy as a function of the $Q_{ij}$ and $P_{ij}$.
At the saddle points we expect to find $Q_{ij} = \widetilde{Q}_{ij}$ and $P_{ij} = \widetilde{P}_{ij}$.
However, $F$ serves as a variational estimate even away from the saddle points.

We close this section by expressing the $Q_{ij}$ and $P_{ij}$ in terms of the momentum-space order
parameters used in Ref.~\onlinecite{rolando}. For the superconducting order, we have
\beq
Q_{ij} = \frac{1}{V} \sum_{\bk} e^{i \bk \cdot ( \br_i - \br_j ) } \Delta_S (\bk),
\eeq
where $V$ is the system volume. The $d$-wave superconductor corresponds to $\Delta_S (\bk) = 2 \Delta_0 (\cos (k_x) - \cos(k_y))$, which 
implies
\bea
Q_{i,i+\hat{x}} &=& \Delta_0 \nn 
Q_{i,i+\hat{y}} &=& -\Delta_0 
\eea
with $\Delta_S$ real (without loss of generality). 

For the charge order, we have a set of orders $\Delta_\bH (\bk)$ for ordering wavevectors $\bH$, and these are related to the
$P_{ij}$ via
\beq
P_{ij} = \sum_{\bH} \left[\frac{1}{V} \sum_{\bk} e^{i \bk \cdot ( \br_i - \br_j ) } \Delta_\bH (\bk) \right] e^{i \bH \cdot ( \br_i + \br_j )/2}.
\eeq
The momentum space charge orders must obey $\Delta_{-\bH} (\bk) = \Delta_{\bH}^\ast (\bk)$.
And solutions which preserve time-reversal symmetry have $\Delta_{\bH} (\bk) = \Delta_{\bH} (-\bk)$.
In our models with bond order parameters only along nearest neighbor links, we must have
\bea
&& \Delta_\bH (\bk) = \\
&& \quad C_1 \cos (k_x) + C_2 \cos (k_y) + C_3 \sin(k_x) + C_4 \sin (k_y) \nonumber
\eea
where $C_{1-4}$ are $\bH$-dependent complex numbers. For the optimum incommensurate charge density wave state 
of Ref.~\onlinecite{rolando}
at wavevector $\bH \approx (H_0,H_0)$, we have $C_2=-C_1$ and $C_3=C_4=0$.
Similarly, for the `staggered-flux' state of Refs.~\onlinecite{marston,kotliar,sudip,leewen,laughlin} we have $\bH = (\pi, \pi)$, 
$C_4=-C_3$ and  $C_1=C_2=0$. And the current-carrying states of Ref.~\onlinecite{varma} have
$\bH = (0,0)$, $C_1=C_2=0$, and $C_3 \neq 0$, $C_4 \neq 0$.

\section{Numerical solutions}
\label{sec:num}

The mean-field ground state of the Hamiltonian in Eq.~\ref{eq:H} can be obtained by minimizing 
the free-energy $F[P,Q]$ defined in  Eq.~\ref{eq:F} relative to the mean-field BDW potentials $P_{ij}$
and the SC pair potentials $Q_{ij}$.
As described in Appendix B, the first derivatives of $F$ in the potentials $P_{ij}$ and $Q_{ij}$ 
i.e. $\partial_{P_{ij}}F$ and 
$\partial_{Q_{ij}}F$, can be computed using perturbation theory. 
Local minima of the free-energy $F[P_{ij},Q_{ij}]$ in the variables $P_{ij}$ and $Q_{ij}$, which 
are then solutions of the set of equations $\partial_{P}F=\partial_Q F=0$, are
 obtained  by solving 
the minimization problem using the Quasi-Newton method as implemented in 
 the fminunc routine in MATLAB. The Quasi-Newton iterations are continued until the derivatives 
of the free-energy $F$ fall below $10^{-5}$. We verify that the obtained solution is the
 global minimum by choosing the initial values of $P$ and $Q$ that we start the minimization from. 
For a true global minima, general values of the initial perturbations $P_{ij}$ and $Q_{ij}$ lead to 
the same final ground state solution at the end of the iterations. 
It is possible to constrain the symmetry of the solution obtained by constraining the 
initial values of $P_{ij}$ and $Q_{ij}$. Therefore, if we choose $P=Q=0$ in the initial state, 
the final solution will continue to respect translational, $4$-fold rotational and $U(1)$ symmetry
 and will not develop either BDW or SC order. It however can develop a uniform value $P_{ij}=P^{(0)}$ 
proportional to the nearest neighbor repulsion $V$. 
Since the final solution will continues to obey time-reversal symmetry, $P^{(0)}$ is found to be real. 
The hopping parameter $t_1$ in the microscopic Hamiltonian in Eq.~\ref{eq:H} is a phenomenological parameter, 
which is chosen to reproduce the bandwidth of the electron dispersion that is measured in experiment. 
We therefore subtract $P^{(0)}$ from the bare hopping so that $t_1+P^{(0)}$ becomes equal to the measured hopping 
parameter.
 
For our calculations we choose an $N\times N$ lattice site unit cell with $N=24$ so that any translation symmetry breaking 
that we obtain as a result of interaction must be commensurate with  the $24\times 24$ unit cell. 
We find our results to be qualitatively similar for larger values of $N$. While the Hamiltonian is periodic with the $N$ 
lattice sites in each direction, the electronic eigenvectors $\Psi(\bm r)$ (where $\bm r$ are positions on the lattice 
sites) described in Eq.~\ref{eq:ev} of the mean-field Hamiltonian 
obey phase twisted periodic boundary conditions $\Psi_{\bm q}(\bm r+\bm R)=e^{i\bm q\cdot\bm R}\Psi_{\bm q}(\bm r)$, where $\bm q$ is a
 Bloch wave-vector
and $\bm R$ is either $\bm R=N\hat{\bm x}$ or  $\bm R=N\hat{\bm y}$. Thus eigenstates in our periodic lattice are 
indexed by the Bloch wave-vector $\bm q=2\pi(n_x \hat{\bm x}+n_y \hat{\bm y})/M$, 
where $n_x$ and $n_y$ are integers from $0$ to $M-1$. For our calculation we choose $M=4$, so that the electronic states 
effectively correspond to periodic boundary conditions on a supercell which is $N M=96$ lattice sites in each direction.  
The value of $M$ is chosen not only to converge the final answer obtained, but also to suppress translational symmetry 
breaking within the unit cell that is induced by the magnetic field. While the magnetic field, in the absence of 
superconductivity does not break translational invariance, the finite size effects resulting from a finite grid 
of Bloch vectors lead to a translational symmetry breaking similar to the way a uniform magnetic flux on a torus breaks 
translational invariance. Our choice of $M=4$ produces a negligible translational symmetry breaking effect even in the 
presence of a magnetic flux.

Following previous work \cite{rolando} we observe that the free-energy $F$ at infinitesimal strength
 of symmetry breaking 
in the BDW and SC sectors  can be analyzed by expanding the free-energy
 $F[P_{ij},Q_{ij},J_{ij},V_{ij}]$ to quadratic order in $P$ and $Q$.
 For a critical strengths of the interaction parameters $J_{ij}$ and $V_{ij}$, the state with $P_{ij}=Q_{ij}=0$ 
is unstable towards a symmetry broken state with either BDW order (i.e. $P_{ij}\neq 0$) or SC order (i.e. with $Q_{ij}\neq 0$).
 In this limit $F$ can be decomposed as 
\beq
F[P_{ij},Q_{ij},J_{ij},V_{ij}]=F_{BDW}[P_{ij},J^{BDW}_{ij}]+F_{SC}[Q_{ij},J^{SC}_{ij}],
\eeq 
where 
\beq 
J^{BDW}_{ij}=J_{ij}+4 V_{ij}/3
\eeq
 parametrizes the instability to symmetry breaking in the BDW channel and 
\beq 
J^{SC}_{ij}=J_{ij}-4 V_{ij}/3
\eeq
parameterizes the instability in the SC channel.

\begin{figure}
\centering
\includegraphics[width=3.4in]{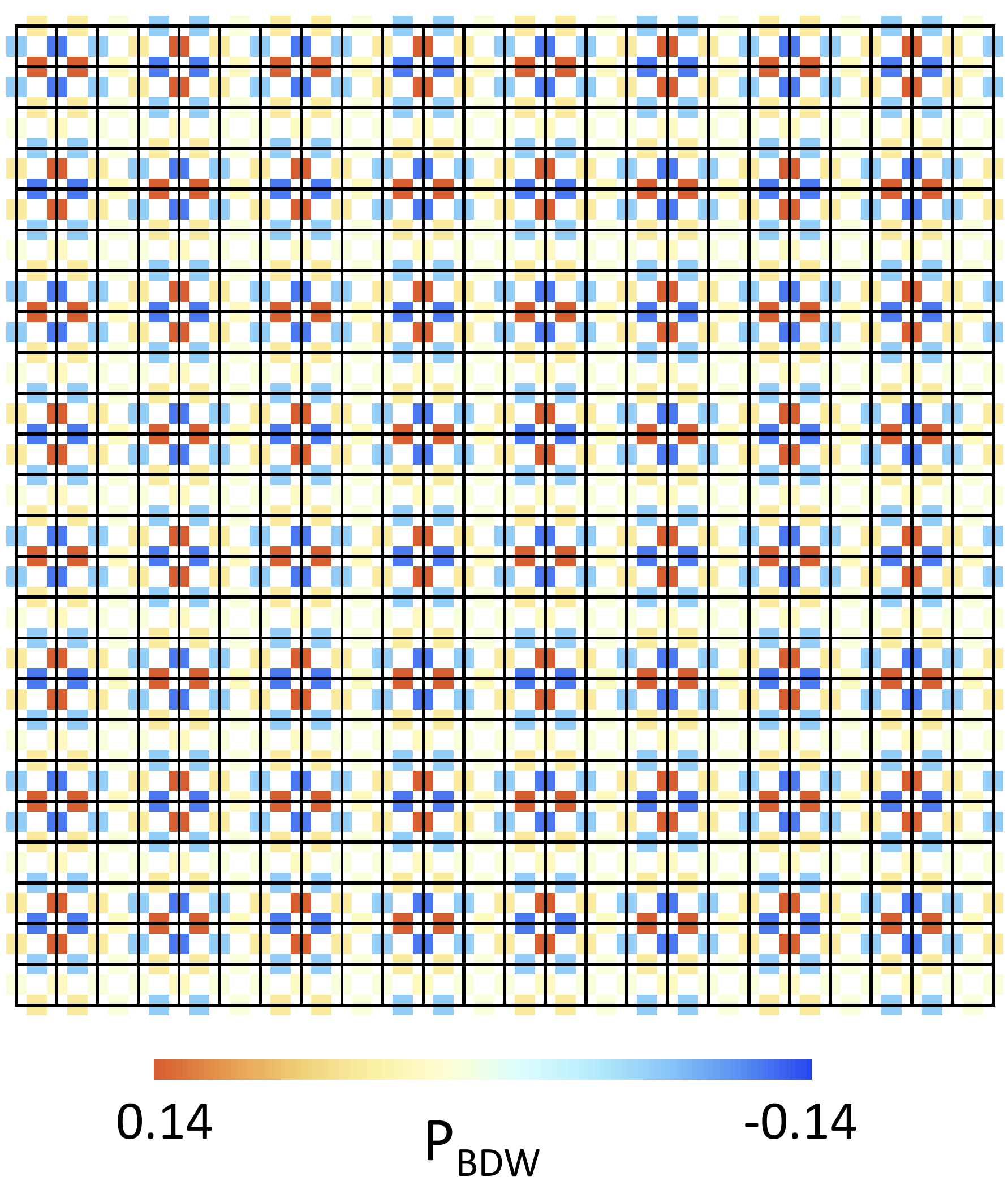}
\caption{BDW order parameter plotted on the $24\times 24$ unit cell at $J^{SC}=0$ and $J^{BDW}=2.4$ (defined in the text) 
in the absence of a B-field. The lines intersect on the lattice (Cu) sites, while the colored squares on the bonds (O sites) represent
the values of $P_{ij}$ on each bond.
By Fourier transforming, we find that the BDW contains Fourier components at both $(\pi/3,\pm \pi/3)$,
whose magnitude and direction essentially coincide with those found in previous work \cite{rolando}. Our calculation shows that the 
checker-board pattern with both wave-vectors co-existing is the energetically favored result. Repeating the calculation at a finite 
magnetic field with one electronic flux quantum per unit cell does not significantly alter this pattern.}
\label{fig:BDW1}
\end{figure}

By minimizing the free-energy for the value $J^{SC}=0$ for different values of $J^{BDW}$,
 we find that the non-interacting ground state is 
unstable towards the formation of a BDW of the form shown in Fig.~\ref{fig:BDW1} for $J^{BDW}\geq 2.4$ 
at a temperature $T=0.1$. Alternatively, it is possible to study the formation of the BDW state 
without breaking the superconducting $U(1)$ symmetry by setting the initial value of $Q=0$.
Note that the ordered state is essentially that found in the quadratic instability computation of Ref.~\onlinecite{rolando}:
it is a bond-ordered state with bi-directional order at the
wavevectors $\approx (\pm H_0, \pm H_0)$ and an internal $d$-wave angular momentum
for the particle-hole pair.
Next, we introduce a magnetic field, as mentioned previously, with one electronic flux quantum 
per the $24\times 24$ lattice: this corresponds to a relatively large field of $\sim 14 T$.
 As 
described at the end of Sec. II and Appendix A, the magnetic 
field is introduced by adding a complex phase to the hopping parameters in Eq.~\ref{eq:H} that 
corresponds to the vector potential from the magnetic field.   Even at this 
relatively large field we find a negligible effect of the magnetic field on the wave-vector of the BDW order 
parameter.

On introducing a finite value for the superconducting coupling constant $J^{SC}\neq 0$ in the absence of a 
magnetic field one expects a finite value of the superconducting order parameter. 
Since the BDW does not gap the entire Fermi surface and the superconducting instability is guaranteed 
to occur for attractive interactions at sufficiently low temperatures, one expects to find SC coinciding with BDW 
at small $J^{SC}$ at low temperatures. The presence of the BDW reduces the superconducting transition temperature 
and the order parameter, but does not eliminate the SC state completely. This indicates a competition between the two 
phases.

\begin{figure}[H]
\centering
\includegraphics[width=3.1in]{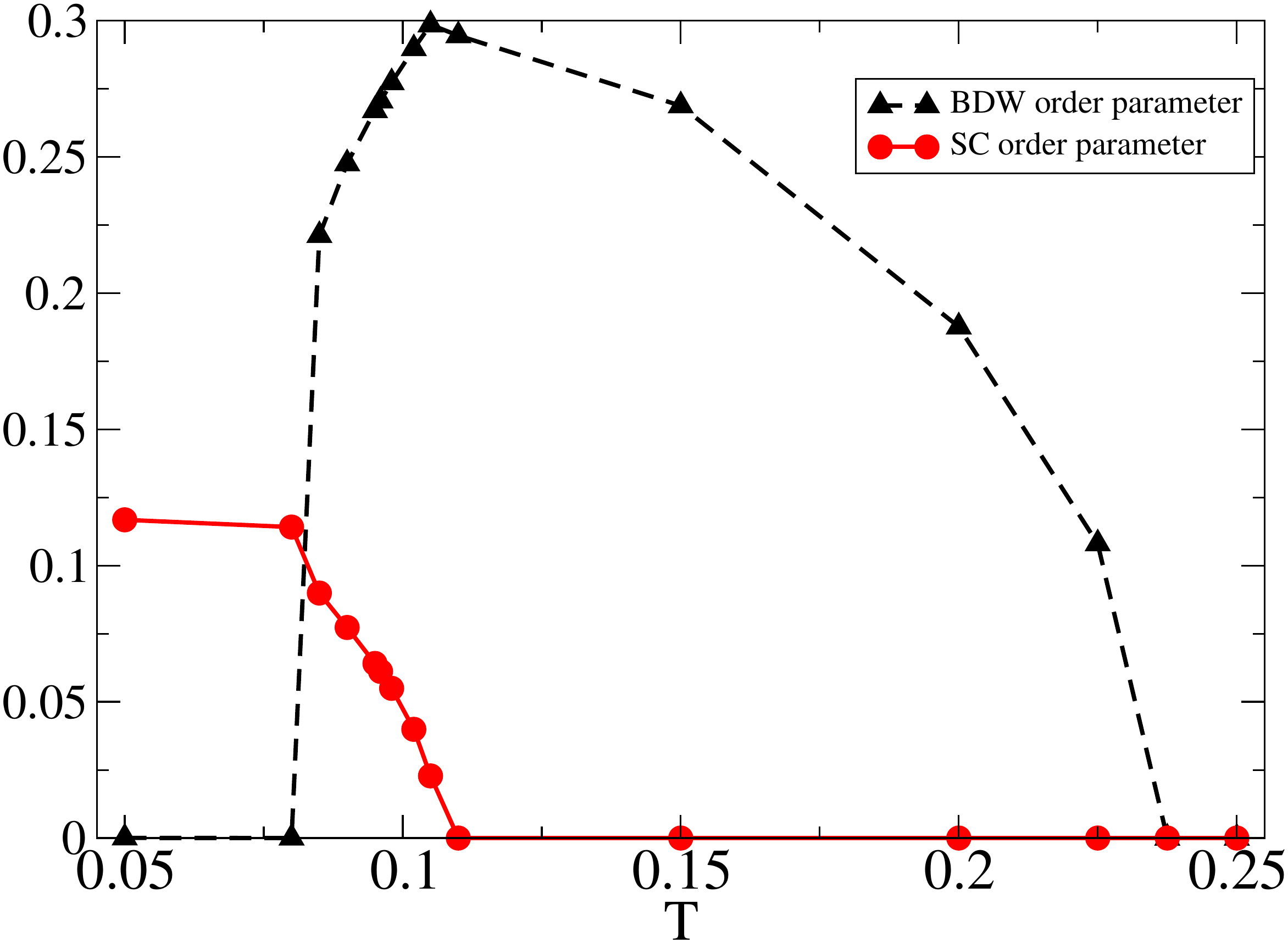}
\caption{
BDW and SC order parameter as a function of temperature $T$ (in units where $t_1=1$) demonstrates coexistence and 
compitetion between the order parameters. For this plot we choose $J^{SC}=1.8$ and $J^{BDW}=2.8$.  
}
\label{fig:BDWSC}
\end{figure}

\begin{figure}[H]
\centering
\includegraphics[width=3.5in]{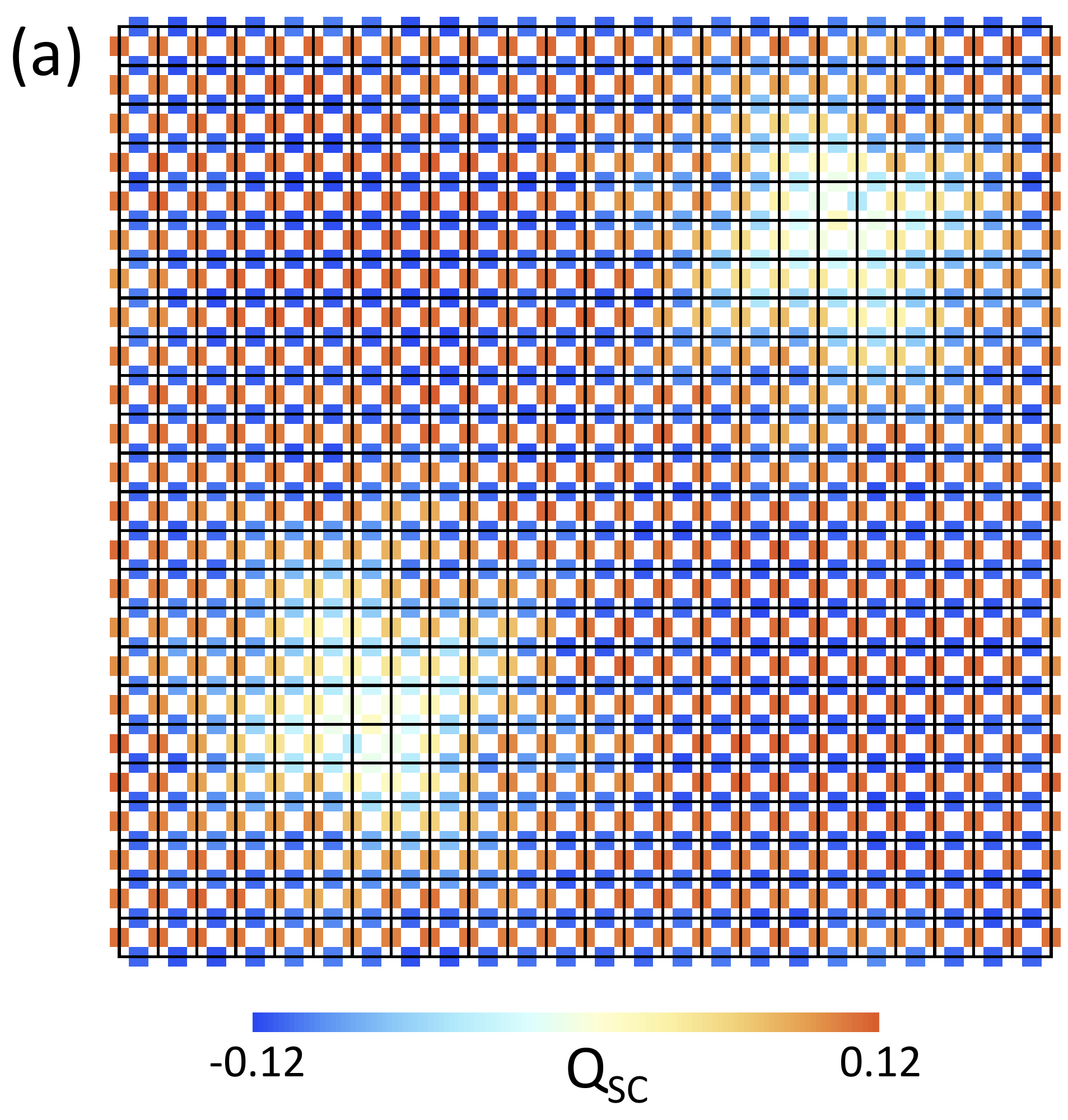}\\~\\~\\
\includegraphics[width=3.5in]{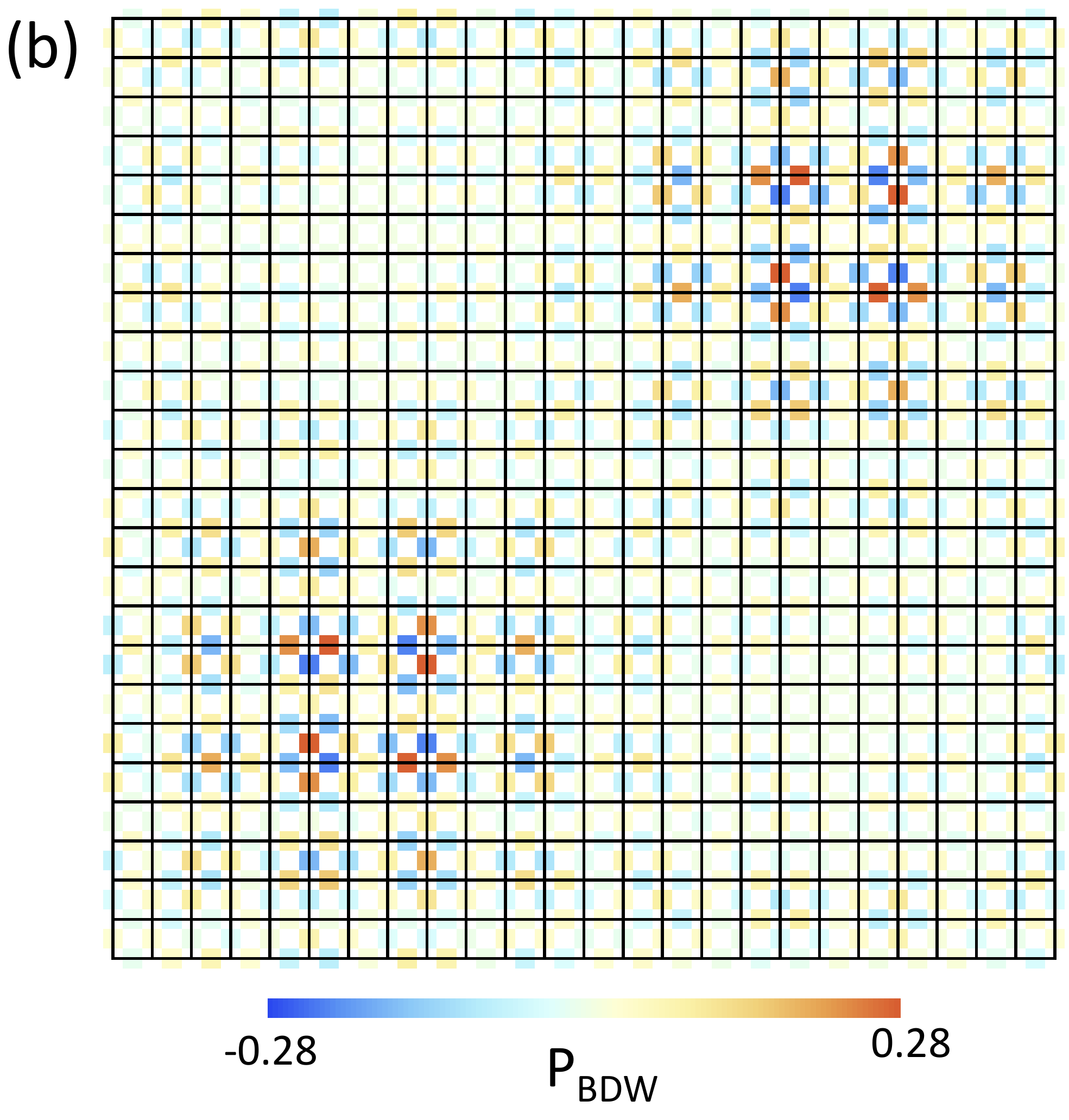}
\caption{
(a) Real part of superconducting order parameter in the presence of a magnetic field. The SC order 
parameter shows the $d$-wave symmetry as a difference of signs between horizontal and vertical bonds. 
The SC order parameter is suppressed near the vortex core. (b) The suppression of the SC order 
parameter allows the nucleation of a BDW halo near the vortex core. For this plot we have chosen 
$J^{SC}=1.9$ and $J^{BDW}=2.8$ and $T=0.09$, which is near the coexistence region. 
}
\label{fig:halo}
\end{figure}

The competition between BDW and SC is seen clearly from Fig.~\ref{fig:BDWSC}. 
At sufficiently high $T>0.11$, the SC order parameter vanishes even though in the range of $T$ considered here, the 
BDW order parameter with a structure similar to that shown in Fig.~\ref{fig:BDW1} exists. As one lowers the temperature 
, in the presence of BDW, one sees the onset of a SC order parameter below $T<0.11$. 
As expected from the exchange 
interaction $J_{ij}$ the superconductivity is of $d$-wave symmetry as is manifest from the difference between the  
signs of the order parameters on the horizontal and vertical bonds in Fig.~\ref{fig:halo}(a). 
In the intermediate regime of temperature $0.085<T<0.11$, for the parameters in Fig.~\ref{fig:BDWSC},
 one finds coexistence  between  BDW and SC order parameters.
 On the other hand 
as  $T$ is reduced further below $T<0.085$ the SC amplitude increases sufficiently so that it dominates 
over the BDW order. At this point, the coexistant state between the SC and BDW becomes energetically unfavorable 
compared to the pure SC state and the BDW order parameter jumps to zero in a transition that is first order within 
mean-field theory. At the same 
temperature $T\sim 0.085$ the SC also increases discontinously.


The application of a magnetic field leads to interesting behavior in the vicinity of the coexistence 
region of the BDW and SC order parameters. The magnetic field is introduced using a vector potential, 
as discussed at the end of Sec. II, and leads to a pair of vortex cores per unit cell where the 
SC order parameter is suppressed as seen in Fig.~\ref{fig:halo}(a). As seen in Fig.~\ref{fig:halo}(b), 
the suppression of the SC order parameter near the vortex core favors the nucleation of a BDW halo 
near the vortex core. This is expected from Fig.~\ref{fig:BDWSC} where we see that a large SC order 
parameter destabilizes the BDW order parameter at low temperatures. Similarly, one might expect that the 
large SC order parameter far from the vortex core might destabilize the BDW state.

\section{Hot spot model}
\label{sec:hotspot}

Our solution of the full lattice model in Section~\ref{sec:num} only found stable charge order at the wavevectors
$(H_0, \pm H_0)$ defined from the hot spots in Fig.~\ref{fig:bz}. This suggests that we may be able extract similar physics
in a model which focuses only on the vicinity of the hotspots. We will propose such a model here, and show that it allows rapid
computation of the equilibrium phase diagram; the model has similarities to the structure of earlier studies of the competition
between commensurate spin density wave order and superconductivity\cite{rafael}. However, our 
model cannot be extended to spatially inhomogeneous situations, as it is
defined in momentum space, and introduces an artificial sharp cutoff at the end of Fermi arcs. 

\begin{figure}
\centering
\includegraphics[width=2.8in]{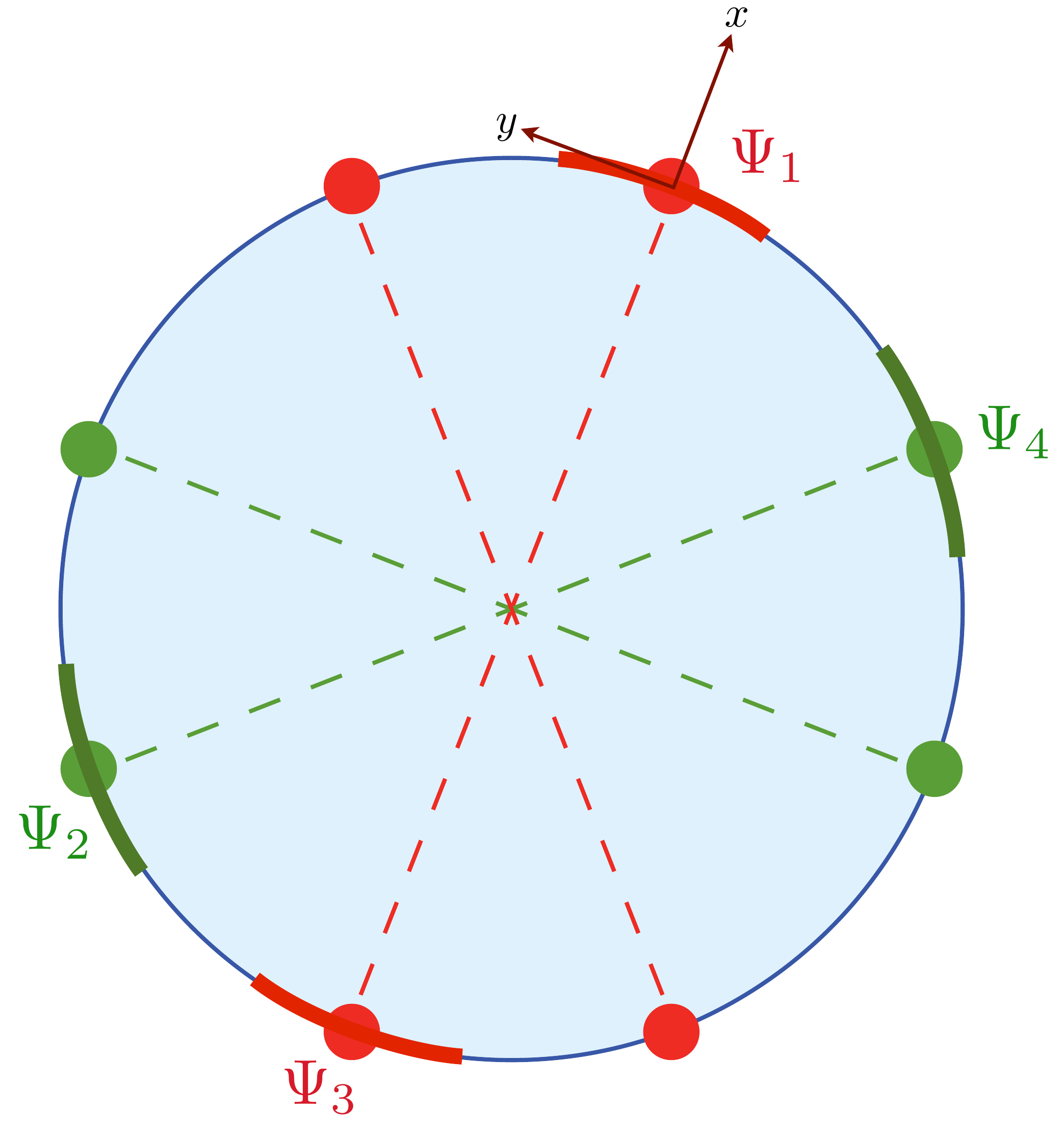}
\caption{Definitions of the $\Psi_{1,2,3,4}$ fermions around the Fermi surface. Each fermion resides around a curved patch
of the Fermi surface shown by the thick lines. The Fermi surface is centered at the corners of the Brillouin zone shown in Fig.~\ref{fig:bz}.
The red (green) hot spots are where the superconducting and bond density wave orders are positive (negative).}
\label{fig:hotspots}
\end{figure}

We introduce 4 species of fermions $\Psi_{a \alpha}$, $a = 1 \ldots 4$, $\alpha = \uparrow, \downarrow$ which reside in the vicinities
of 4 of the hot spots, as shown in Fig.~\ref{fig:hotspots}. 
Their dispersions are defined by the momentum space theory
\bea
&& H_0 =  \sum_{\bk} \Biggl[ \epsilon_1 (\bk)  \, \Psi_{1 \alpha}^\dagger (\bk)  \Psi_{1 \alpha}^{\vphantom \dagger} (\bk)
+ \epsilon_2 (\bk)\, \Psi_{2 \alpha}^\dagger (\bk)  \Psi_{2 \alpha}^{\vphantom \dagger} (\bk) +\nn
&& \epsilon_1 (- \bk) \, \Psi_{3 \alpha}^\dagger (\bk)  \Psi_{3 \alpha}^{\vphantom \dagger} (\bk)
+ \epsilon_2 (- \bk) \, \Psi_{4 \alpha}^\dagger (\bk)  \Psi_{4 \alpha}^{\vphantom \dagger} (\bk) \Biggr].
\eea
We take the origin of momentum space at the hot spots, and orient the $x$-axis orthogonal to the Fermi surface
for the $\Psi_{1,3}$ fermions; so we can write
\beq
\epsilon_1 (\bk) = k_x + \gamma k_y^2 . \label{eps1}
\eeq
We have taken the Fermi velocity to be unit, while $\gamma$ measures the curvature of the Fermi surface. 
The dispersion $\epsilon_2 (\bk)$ has the form obtained by rotating $\epsilon_1 (\bk)$ so that the direction
orthogonal to the Fermi surfaces of the $\Psi_{2,4}$ has a linear dispersion; we will not need its explicit form
and so do not write it out.
We chose the convenient momentum space 
cutoffs $- \Lambda < k_x, k_y < \Lambda$, and $\sum_\bk \equiv \int d^2 k/\Lambda^2$, and the value $\gamma = 1/\Lambda$. 
Here $\Lambda \ll \pi$ in the units of the underlying lattice, so that we are only accounting for the immediate vicinities of the hot spots.
However, by rescaling momenta and all the couplings in our continuum model we can change the value $\Lambda$, and we use units
in which $\Lambda = \pi$.

\begin{figure}
\centering
\includegraphics[width=1.3in]{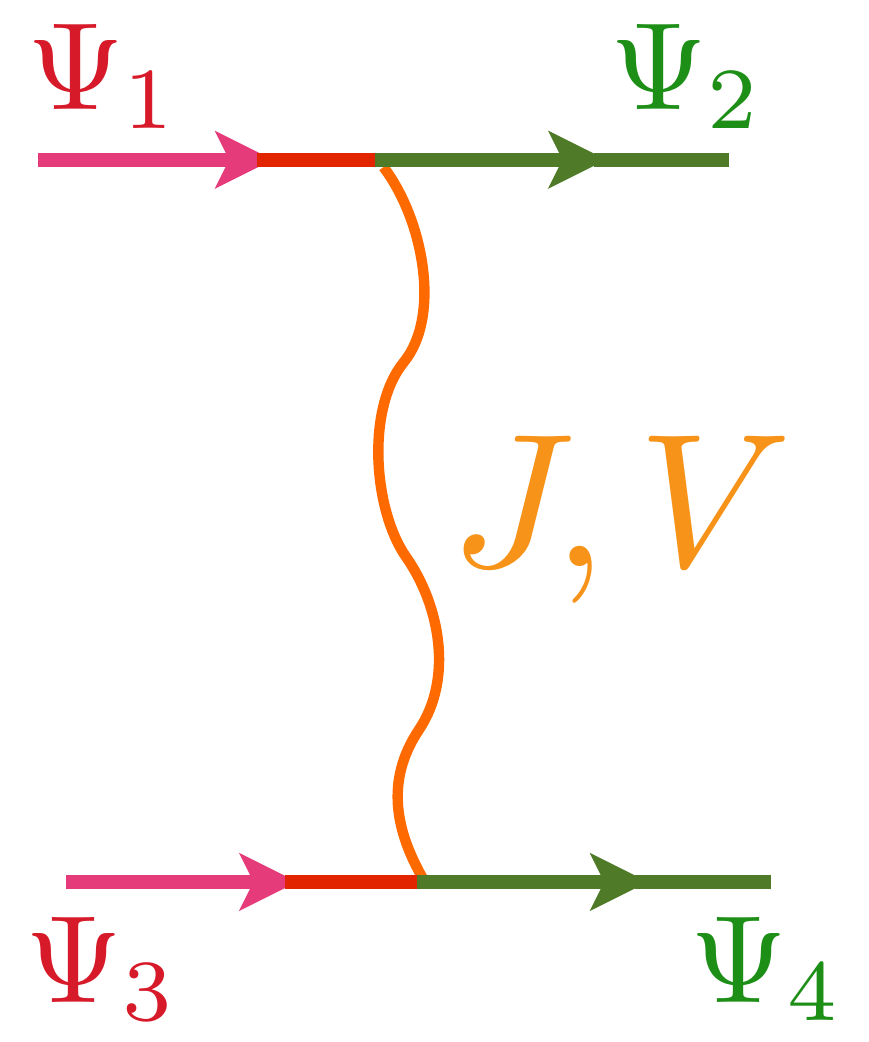}
\caption{Interactions between the $\Psi_{1,2,3,4}$ fermions.}
\label{fig:JV}
\end{figure}

Next, we add interactions between these fermions. For this, we simply include the $J$ and $V$ terms of the lattice model,
and project out the terms which lead to scattering between the hotspots, as illustrated in Fig.~\ref{fig:JV}. 
This gives us
\bea
H_1 &=& \int d^2 x \Biggl[
 - J \left(  \Psi_{1 \alpha}^\dagger \vec{\sigma}_{\alpha\beta} \Psi_{2 \beta}^{\vphantom \dagger} 
+  \Psi_{2 \alpha}^\dagger \vec{\sigma}_{\alpha\beta} \Psi_{1 \beta}^{\vphantom \dagger} \right)  \nn
&~& \quad\quad\quad\quad \cdot
\left(  \Psi_{3 \gamma}^\dagger \vec{\sigma}_{\gamma\delta} \Psi_{4 \delta}^{\vphantom \dagger} 
 + \Psi_{4 \gamma}^\dagger \vec{\sigma}_{\gamma\delta} \Psi_{3 \delta}^{\vphantom \dagger}  \right) \\
&~&  - V \left(  \Psi_{1 \alpha}^\dagger \Psi_{2 \alpha}^{\vphantom \dagger} 
+  \Psi_{2 \alpha}^\dagger  \Psi_{1 \alpha}^{\vphantom \dagger} \right)  
\left(  \Psi_{3 \beta}^\dagger \Psi_{4 \beta}^{\vphantom \dagger} 
 + \Psi_{4 \beta}^\dagger  \Psi_{3 \beta}^{\vphantom \dagger}  \right) \Biggr] \nonumber
\eea

The full Hamiltonian $H_0+H_1$ has an exact SU(2)$\times$SU(2) pseudospin rotation symmetry \cite{metlitski10-2} when $\gamma=0$
(so that there is no Fermi surface curvature) and $V=0$.

\begin{figure}[h]
\includegraphics[width=3.5in]{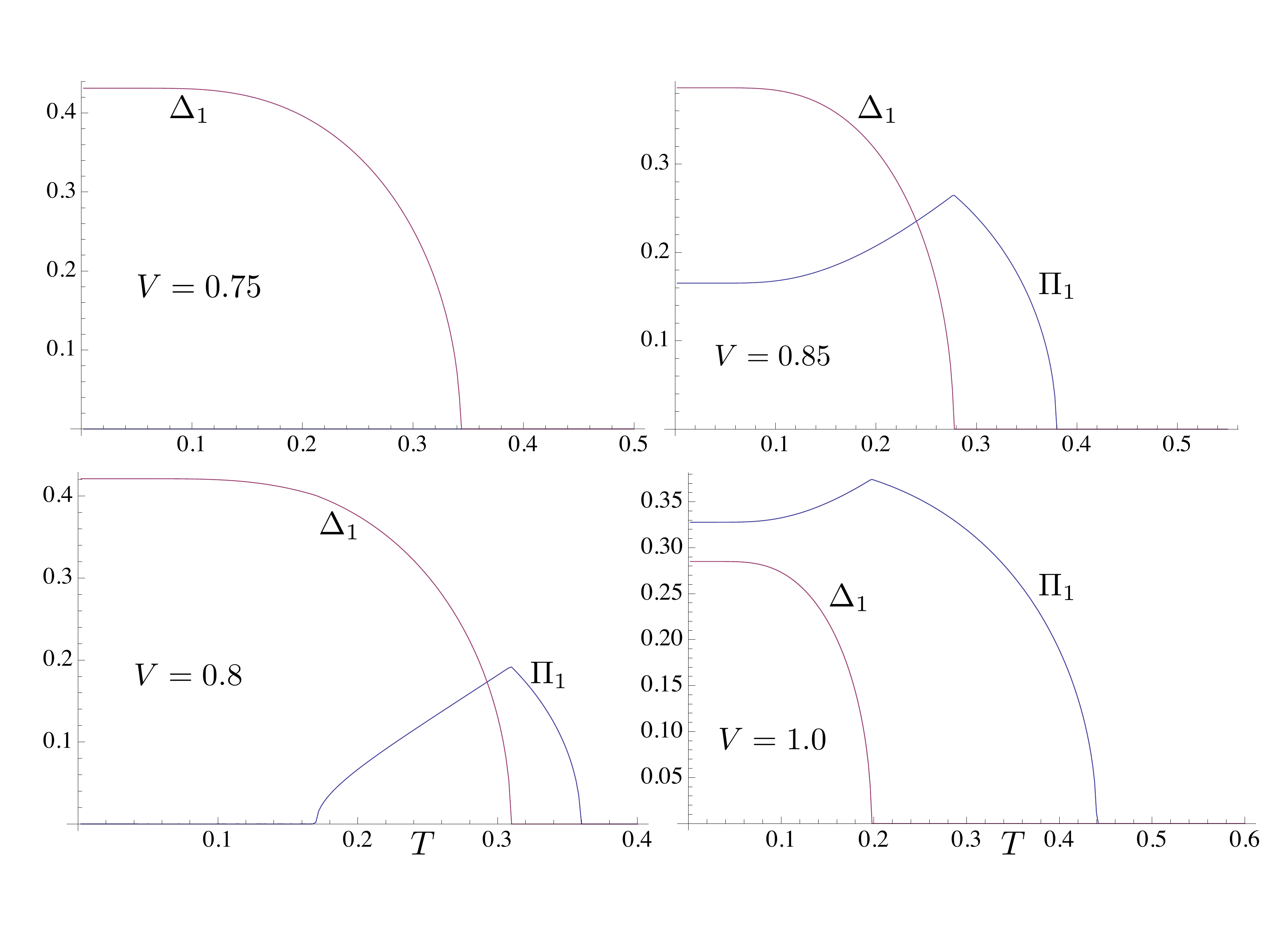}
\caption{Superconducting ($\Delta_1$) and bond ($\Pi_1$) orders in the hot spot model as a function of $T$ and $V$
for $J=1.2$.}
\label{fig:phase}\end{figure}

We now proceed with the Hartree-Fock-BCS theory of the hotspot model $H_0 + H_1$. This is now straightforward because it is easy to identify
the bond density wave order parameter as the particle-hole pair condensate of fermions on antipodal points of the Fermi surface. 
We therefore introduce the condensates
\bea
\Delta_1 (\bk) &=& \left\langle \varepsilon_{\alpha\beta} \Psi_{1\alpha}^\dagger (\bk) \Psi_{3\beta}^{\dagger} (-\bk) \right\rangle 
\quad;\quad \Delta_1 \equiv \sum_\bk \Delta_1 (\bk) \nn
\Delta_2 (\bk) &=& \left\langle \varepsilon_{\alpha\beta} \Psi_{2\alpha}^\dagger (\bk) \Psi_{4\beta}^{\dagger} (-\bk) \right\rangle 
\quad;\quad \Delta_2 \equiv \sum_\bk \Delta_2 (\bk)\nn
 \Pi_1 (\bk) &=& \left\langle  \Psi_{1\alpha}^\dagger (\bk) \Psi_{3\alpha}^{\vphantom \dagger} (\bk) \right\rangle
 \quad;\quad \Pi_1 \equiv \sum_\bk \Pi_1 (\bk)\nn
 \Pi_2 (\bk) &=& \left\langle  \Psi_{2\alpha}^\dagger (\bk) \Psi_{4\alpha}^{\vphantom \dagger} (\bk) \right\rangle
\quad;\quad \Pi_2 \equiv \sum_\bk \Pi_2 (\bk) \label{orders}
\eea
The superconducting order parameters are $\Delta_{1,2}$, while the bond density wave order parameters are $\Pi_{1,2}$. 
We will find that optimal state has a $d$-wave signature for both the superconducting and bond orders, with 
$\Delta_1 = - \Delta_2$ and $\Pi_1 = - \Pi_2$. With the above orders, 
the mean field Hamiltonian is
\bea
&& H_{MF} = H_0 +\frac{(3J-V)}{2} \left( - \Delta_1 \, \varepsilon_{\alpha\beta} \Psi_{2\alpha}^{\vphantom \dagger} (\bk) \Psi_{4\beta}^{\vphantom \dagger} (-\bk) \right. \nn
&& + \Delta_2^\ast \, \varepsilon_{\alpha\beta} \Psi_{1\alpha}^{ \dagger} (\bk) \Psi_{3\beta}^{ \dagger} (-\bk) 
 - \Delta_2 \, \varepsilon_{\alpha\beta} \Psi_{1\alpha}^{\vphantom \dagger} (\bk) \Psi_{3\beta}^{\vphantom \dagger} (-\bk) \nn
&& \left. + \Delta_1^\ast \, \varepsilon_{\alpha\beta} \Psi_{2\alpha}^{ \dagger} (\bk) \Psi_{4\beta}^{ \dagger} (-\bk) 
 \right) \nn
 && + \frac{(3J+V)}{2} \left( \Pi_1 \, \Psi_{4 \alpha}^\dagger (\bk) \Psi_{2 \alpha}^{\vphantom \dagger} (\bk) 
 + \Pi_2^\ast \, \Psi_{1 \alpha}^\dagger (\bk) \Psi_{3 \alpha}^{\vphantom \dagger} (\bk) \right. \nn
&& \left. + \Pi_2 \, \Psi_{3 \alpha}^\dagger (\bk) \Psi_{1 \alpha}^{\vphantom \dagger} (\bk) 
 + \Pi_1^\ast \, \Psi_{2 \alpha}^\dagger (\bk) \Psi_{4 \alpha}^{\vphantom \dagger} (\bk) 
 \right) .
\eea
We diagonalize this Hamiltonian by writing the Hamiltonian for $\Psi_{1,3}$ as
\bea
H_{MF} &=& \sum_{\bk} \left( \Psi_{1\uparrow}^\dagger (\bk), \Psi_{3\uparrow}^\dagger (\bk), \Psi_{1\downarrow}^{\vphantom \dagger} (-\bk),
\Psi_{3\downarrow}^{\vphantom \dagger} (-\bk) \right) M(\bk) \nn
&~& \times \left( \begin{array}{c} 
\Psi_{1\uparrow}^{\vphantom \dagger} (\bk) \\ \Psi_{3\uparrow}^{\vphantom \dagger} (\bk) \\ 
\Psi_{1\downarrow}^{ \dagger} (-\bk) \\
\Psi_{3\downarrow}^{ \dagger} (-\bk) \end{array} \right) + \sum_\bk \left( \epsilon_1 (-\bk) + \epsilon_1 (\bk) \right),
\eea
where the $4 \times 4$ matrix $M(\bk)$ is
\bea
M(\bk) &=& \left( 
\begin{array}{cc} 
\epsilon_1 (\bk) & (3J+V) \Pi_2^\ast /2  \\
(3J+V) \Pi_2 /2& \epsilon_1 (-\bk)  \\
0 & (3J-V) \Delta_2 /2 \\
(3J-V) \Delta_2 /2 & 0  \end{array}
\right. \nn
&~& 
\left.
\begin{array}{cc} 
 0  & (3J-V) \Delta_2^\ast /2 \\
 (3J-V) \Delta_2^\ast /2 & 0 \\
-\epsilon_1 (-\bk) & - (3J+V) \Pi_2 /2\\
 - (3 J + V) \Pi_2^\ast /2& - \epsilon_1 (\bk) \end{array}
\right).
\eea
Let $U (\bk)$ be the unitary matrix which diagonalizes $M (\bk)$:
\beq
U^\dagger (\bk) M (\bk) U (\bk) = \Lambda (\bk) 
\eeq
where $\Lambda (\bk) $ is a diagonal matrix with entries $\lambda_i (\bk) $. Then
\bea
\sum_\bk \left\langle \Psi_{1 \uparrow}^\dagger (\bk) \Psi_{3 \uparrow}^{\vphantom \dagger} (\bk) \right\rangle &=& \sum_{\bk} \sum_i  U^\ast_{1i} (\bk) U_{2i} (\bk ) f(\lambda_i (\bk) ) \nn & \equiv & {\widetilde{\Pi}_1}/{2} \nn
\sum_\bk \left\langle \Psi_{1 \downarrow}^\dagger (-\bk) \Psi_{3 \downarrow}^{\vphantom \dagger} (-\bk) \right\rangle &=& -\sum_{\bk} \sum_i  U_{3i} (\bk) U^\ast_{4i} (\bk ) f(\lambda_i (\bk) ) \nn & \equiv&  {\widetilde{\Pi}_1}/{2} \nn
\sum_\bk \left\langle \Psi_{1 \uparrow}^\dagger (\bk) \Psi_{3 \downarrow}^{ \dagger} (-\bk) \right\rangle &=& \sum_{\bk} \sum_i  U^\ast_{1i} (\bk) U_{4i} (\bk ) f(\lambda_i (\bk) ) \nn &\equiv&  {\widetilde{\Delta}_1}/{2} \nn
\sum_\bk \left\langle \Psi_{1 \downarrow}^\dagger (-\bk) \Psi_{3 \uparrow}^{ \dagger} (\bk) \right\rangle &=& -\sum_{\bk} \sum_i  U_{3i} (\bk) U^\ast_{2i} (\bk ) f(\lambda_i (\bk) )  \nn &\equiv& - {\widetilde{\Delta}_1}/{2} 
\eea
Assuming the state with $\Delta_1=-\Delta_2$ and $\Pi_1=-\Pi_2$, the free energy is
\bea
\frac{F}{2} &=& F_{MF} + \left\langle \frac{H}{2} - H_{MF} \right\rangle_{MF} \nn
&=& \sum_\bk \left( \epsilon_1 (-\bk) + \epsilon_1 (\bk) \right) - T \sum_\bk \sum_i \ln \left( 1 + e^{-\lambda_i (\bk)/T} \right) 
\nn &+& \frac{(3J-V)}{2} \left( - \widetilde{\Delta}_1^\ast \widetilde{\Delta}_1 + \Delta_1^\ast \widetilde{\Delta}_1 + \widetilde{\Delta}_1^\ast \Delta_1 \right)\nn
&+& \frac{(3J+V)}{2} \left( - \widetilde{\Pi}_1^\ast \widetilde{\Pi}_1 + \Pi_1^\ast \widetilde{\Pi}_1 + \widetilde{\Pi}_1^\ast \Pi_1 \right).
\eea
We determined the phase diagrams by solving the above mean-field equations for 
$J_1 = 1.2$, and the phase diagrams
as a function of $V$ are in Fig.~\ref{fig:phase}.
Note the similarity of the $T$ evolution in the region of co-existing orders to that obtained in the solution 
of the full lattice model earlier in Fig.~\ref{fig:BDWSC}.

\begin{figure}[H]
\centering
\includegraphics[width=3.4in]{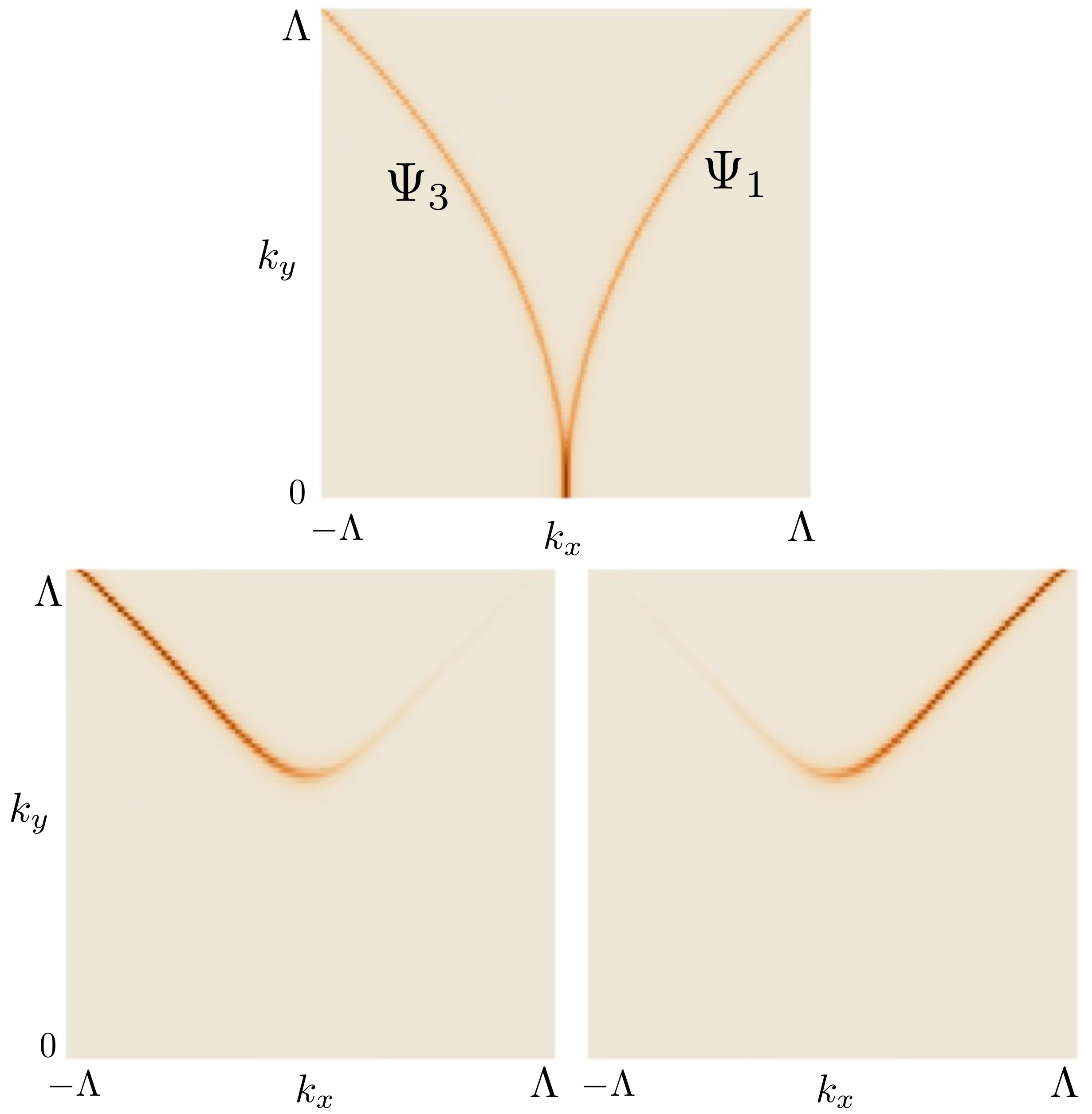}
\caption{Spectral density of the $\Psi_1$ and $\Psi_3$ fermions at zero frequency. The top figure shows both fermions
in the metal with no ordering, and so plots $\epsilon_1 (\pm \bk)$ from Eq.~(\ref{eps1}). 
These Fermi surfaces are those in Fig.~\ref{fig:hotspots}, but shifted in momentum space so that their hotspots coincide.
The displayed momenta are in a small vicinity of the hotspots, with momentum
cutoffs are $\pm \Lambda$.
The bottom figures show the
spectral densities of $\Psi_{1,3}$ respectively in the presence of bond density wave order, $\Pi_1 \neq 0$. Note that this
opens up a gap near $\bk = 0$, but reconstructed Fermi surfaces remain.}
\label{fig:reconstruct}
\end{figure}

The basic features of Figs.~\ref{fig:phase} can be understood by interplay 
between the two terms which break the pseudospin symmetry: the Fermi surface curvature $\gamma$ (which prefers 
superconductivity)
and the nearest-neighbor Coulomb repulsion, $V$ (which prefers bond order). When $\gamma=V=0$, the two orders
are degenerate, and the free energy can be shown to depend only upon $|\Delta_1|^2 + |\Pi_1|^2$. This is a consequence of the pseudospin symmetry, and any co-existence state with the same overall
magnitude is also degenerate. When we turn on only $\gamma$, but keep $V=0$, superconductivity appears first upon lowering $T$; this gaps
out the Fermi surface completely (in the present hotspot model), and bond order never appears down to the lowest $T$. The same situation
remains when a small $V>0$ is turned on, and indeed as long as superconductivity is the first instability upon lowering $T$.
On the other hand, when $V$ is large enough and positive, the first instability upon lowering $T$ is to bond order.
At its initial onset, the bond order only gaps out the Fermi surface in the immediate vicinity of the hot spots, but a reconstructed Fermi surface
does appear; see Fig.~\ref{fig:reconstruct}. 
At lower $T$, this reconstructed Fermi surface undergoes a BCS instability, yielding a state with co-existing superconductivity
and bond order. And as we lower $T$ further, the superconductivity continues to increase in strength at the expense of the bond order: this is because
the superconductivity has the Cooper-logarithm in its susceptibility irrespective of the Fermi surface curvature, while the bond order is suppressed
by the Fermi surface curvature.

\section{Conclusions}
\label{sec:conc}

A shortcoming in the experimental applications of the present mean-field computations of the $t$-$J$-$V$ model
is that they have consistently preferred an incommensurate $d$-wave charge density wave order along the
$(1, \pm 1)$ directions. This is in contrast to previous treatments\cite{vojta1,vojta2,vojta3,vojta4} 
of the $t$-$J$-$V$ model at $T=0$ in the superconducting state,
which also imposed an on-site $U= \infty$ constraint in a particular large $N$ limit; the latter computations found commensurate
$d$-wave bond order, but oriented along the $(1,0)$, $(0,1)$ directions, as in the experimental observations.\cite{keimer,chang,hawthorn}
On the other hand, in these $U=\infty$ computations, the Fermi surface structure appeared to play no role in determining
the magnitude of the ordering wavevector. This suggests to us that it would be worthwhile to examine the instabilities of the
$t$-$J$-$V$ model in the high temperature normal state, while also imposing the $U=\infty$ constraint: such a computation
could determine the mechanism of the orientation of the ordering wavevector, while also displaying the role of the Fermi surface and the hot spots.

Our present results, along with the recent results of Ref.~\onlinecite{o6}, also suggest a recipe for obtaining higher critical
temperatures for superconductivity. The point made in Ref.~\onlinecite{o6} is that it is the combined instabilities of the high
temperature metal to superconductivity and charge order which lead to a large regime of fluctuations in the pseudogap regime.
So we need to suppress the charge ordering instability, while preserving superconductivity. The model of Section~\ref{sec:hotspot}
showed this can be achieved by increasing the Fermi surface curvature. At the same time, we need a large $J$ to maintain
the pairing instability. We note that large Fermi surface curvatures are found in the pnictides, which have so far not shown any
charge-ordering instabilities, as expected in our approach; on the other hand, the larger $J$'s are in the cuprates.
It would therefore be worthwhile to search for quasi-two-dimensional
compounds which combine these desirable features of the existing high temperature
superconductors. 

\acknowledgements

This research was supported by the National Science Foundation under grant DMR-1103860. J.S. acknowledges the 
Harvard Quantum Optics center for support while the author was at Harvard. J.S. would also like to acknowledge the University of Maryland, Condensed Matter theory center and the Joint Quantum institute for start-up support during the final stages.

\appendix

\section{Magnetic field BDW/SC}
In this section we discuss how we choose the lattice vector potential for the Hamiltonian following Wang and Vafek 
 \cite{vafek}.
The magnetic field enters Eq.~(\ref{eq:t}) through the phase $a_{ij}$. We note that the gauge transformation in the Hamiltonian 
takes the form 
\begin{align}
&c^\dagger_{i,\sigma}\rightarrow c^\dagger_{i,\sigma} e^{i\Lambda_i}\nn
&a_{ij}\rightarrow a_{ij}-(\Lambda_i-\Lambda_j).
\end{align}
Neither the orders $P,Q$ defined in the previous sections are invariant under the gauge transformation. However 
\begin{align}
&\mathcal{P}_{ij}=e^{-i a_{ij}}\expect{c^\dagger_{i,\sigma}c_{j,\sigma}}.
\end{align}
Similarly $\tilde{\rho}_{P,i,j}=e^{-i a_{ij}}\rho_{P,i,j}$ is a gauge-invariant version of $\rho_P$.
The superconducting order parameter $Q$ breaks gauge invariance and there is no gauge invariant form of $Q$.

A choice of $a_{ij}$ formally breaks translation invariance since if $\bm A$ were periodic in some unit cell its integral around the boundary 
would have to vanish forcing the flux in the unit cell to be zero. However, translation invariance can be recovered in a 
lattice system by performing a "singular" gauge transformation where one threads one quantum of flux $\Phi_0$ through a single 
plaquette of the lattice. Since this has no effect on lattice electrons, we can choose $\bm A$ to be periodic and satisfy 
\begin{equation}
\bm\nabla\times \bm A=(B-\Phi_0\delta(\bm r))\hat{\bm z}.
\end{equation}
For a lattice system we use the discrete curl to state this as  
\begin{equation}
A_v(\bm r)-A_h(\bm r)+A_h(\bm r+\hat{y})-A_v(\bm r+\hat{x})=\Phi(\bm r),
\end{equation}
where $A_{v,h}(\bm r)$ are the phases associated with the bonds connecting $\bm r\rightarrow\bm r+\hat{y}$ and 
$\bm r\rightarrow\bm r+\hat{x}$ respectively. The flux $\Phi(\bm r)$ is contained in the plaquette surrounded 
by the lattice sites $\bm r,\bm r+\hat{y},\bm r+\hat{x}+\hat{y},\bm r+\hat{x}$ i.e. it is 
 the plaquette containing $\bm r$ at its bottom left corner.

The vector potential $A(\bm r)$ is gauge dependent. For the purpose of representing 
vortices, we choose the gauge which minimizes the magnitude of $A$ i.e. $F=\int d\bm r|\bm\nabla A|^2$.
This gauge has vanishing discrete divergence
\begin{align}
&A_v(\bm r+\hat{y})-A_v(\bm r)+A_h(\bm r+\hat{y})-A_h(\bm r+\hat{y}-\hat{x})=0.
\end{align}
The above divergenceless condition is solved by defining $A$ in terms of plaquette variables $\psi(\bm r)$ (defined 
similar to $\Phi(\bm r)$) so 
that we define 
\begin{align}
&A_v(\bm r+\hat{x})=\psi(\bm r+\hat{x})-\psi(\bm r)\nn
&A_h(\bm r+\hat{y})=-\psi(\bm r+\hat{y})+\psi(\bm r)\nn
&A_v(\bm r+\hat{y})-A_v(\bm r)+A_h(\bm r+\hat{y})-A_h(\bm r+\hat{y}-\hat{x})\nn
&=\psi(\bm r+\hat{y})-\psi(\bm r+\hat{y}-\hat{x})-\psi(\bm r)+\psi(\bm r-\hat{x})\nonumber\\
&-\psi(\bm r+\hat{y})+\psi(\bm r)+\psi(\bm r+\hat{y}-\hat{x})-\psi(\bm r-\hat{x})=0,
\end{align} 
which is now manifestly divergenceless.

Substituting into the curl equation we get  
\begin{align}
&A_v(\bm r)-A_h(\bm r)+A_h(\bm r+\hat{y})-A_v(\bm r+\hat{x})\\
&=4\psi(\bm r)-\psi(\bm r-\hat{x})-\psi(\bm r+\hat{x})-\psi(\bm r+\hat{y})-\psi(\bm r-\hat{y})\nonumber\\
&=\Phi(\bm r).
\end{align}
The above lattice system is easily solved by Fourier transforms by substituting 
\begin{align}
&\psi(\bm r)=\sum_G \psi_G e^{i\bm G\cdot\bm r}\nn
&\Phi(\bm r)=4\sum_{\bm G}\psi_G (\sin^2{G_x/2}+\sin^2{G_y/2})e^{i\bm G\cdot\bm r}\nn
&\Phi(\bm r)=\Phi_0[\delta(\bm r-\bm r_1)+\delta(\bm r-\bm r_2)]-\frac{2\Phi_0}{N_{lat}}\nn
&=\sum_G \Phi_G e^{iG\cdot r}\nn
&\Phi_G=\frac{1}{N_{lat}}\sum_{\bm r}\Phi(\bm r)e^{-i G\cdot\bm r} \nonumber \nn
&~~~~=\frac{\Phi_0}{N_{lat}}[e^{i G\cdot r_1}+e^{i G\cdot r_2}-2\delta_{G=0}]\nn 
&\psi_{G\neq 0}=\frac{\Phi_0}{4 N_{lat}(\sin^2{G_x/2}+\sin^2{G_y/2})}[e^{i G\cdot r_1}+e^{i G\cdot r_2}],
\end{align}
where $\Phi_0=2\pi$.

To minimize the supercurrent, we want to consider a gauge with two spatially separated vortices. This is 
obtained by placing $\Phi_0$ flux in one of the vortices on the diagonal of a square unit cell and having a 
diagonal branch-cut connecting the vortex to the other vortex on the diagonal. The sign of the hopping $t_{jj'}$ 
is flipped along the diagonal branch-cut. While this gauge is arbitrary, the phase of the superconducting order parameter 
is expected to be close to uniform in this gauge.

\section{Evaluating gradients of free-energy $F$}
For the minimization of the free-energy we need to compute the gradient of the free-energy $F$ in Eq.~(\ref{eq:F}).
Writing the order parameter operators $\tilde{P}$ and $\tilde{Q}$ as $\expect{U_n}$ and the 
potential amplitudes as $\lambda_n$, we can write the free-energy in Eq.~\ref{eq:F} in the form 
\begin{align}
&F[\lambda]=F_{MF}[\lambda]=\sum_{n}L_{n,n}\expect{U_n}^2-\lambda_n \expect{U_n}+F_{HF}\nn
&F_{HF}=-T Tr[\textrm{log}(\cosh{\beta H_0})]\nn
&=-T\sum_n -\epsilon_n+\textrm{log}(1+e^{-\beta\epsilon_n(\lambda)})
\end{align} 
where $\epsilon_n$ are eigenvalues of $H_0$.
Using a Taylor expansion in $\lambda$ we note that the first derivative of $F_{HF}$ depends only on the diagonal 
in energy terms of $U$ i.e. 
\begin{align}
&\partial_{\lambda_n}F_{HF}=\sum_p \frac{U_{n,p,p}}{1+e^{\beta \epsilon_p}}=Tr[U_n (1+e^{\beta H_0})^{-1}]\\
&=\expect{U_n}.
\end{align}
Similarly, the second derivative can involve only two states $p,q$ at a time and therefore can be derived by 
considering a $2\times 2$ matrix to be 
\begin{align}
&\partial_{\lambda_n,\lambda_m} F_{HF}=\nn
&~~~-\sum_{p,q} U_{n,p,q}U_{m,q,p} \frac{\textrm{tanh}(\epsilon_p/2 T)-\textrm{tanh}(\epsilon_q/2 T)}{2(\epsilon_p-\epsilon_q)},\label{EqF2}
\end{align}
where one is careful to take the limit $\epsilon_p-\epsilon_q\rightarrow 0$ for the diagonal terms $p=q$.
Using these results 
\begin{align}
&\partial_{\lambda_p} F_{MF}[\lambda]=\sum_n \partial_{\lambda_p}\expect{U_n}(2 L_{n,n}\expect{U_n}-\lambda_n)\\
&=\sum_n (2 L_{n,n}\expect{U_n}-\lambda_n)\partial_{\lambda_n,\lambda_p}F_{HF}.
\end{align}
Using the definiteness of the derivative $\partial_{\lambda_n,\lambda_p}F_{HF}$, we note that the minimum of the free-energy, 
which satisfies $\partial_\lambda F_{MF}=0$ also satisfies the mean field equations 
\begin{align}
&\lambda_n=2 L_{n,n}\expect{U_n}.
\end{align}

\end{document}